\documentclass[review]{elsarticle}

\usepackage[a4paper, margin=2.5cm]{geometry}

\usepackage{lineno,hyperref}
\usepackage{numprint}
\npthousandsep{,}\npthousandthpartsep{}\npdecimalsign{.}
\modulolinenumbers[5]

\journal{Fire Safety Journal}

\usepackage{color}
\usepackage{comment}
\usepackage{enumitem}
\usepackage{tabularx}
\usepackage{array}
\usepackage{color, colortbl}
\usepackage{multirow}
\usepackage{longtable}
\usepackage{pdflscape}
\usepackage{amssymb}

\definecolor{Gray}{gray}{0.9}

\newcommand{\resub}[1]{{#1}}

\definecolor{mygreen}{RGB}{0, 150, 51}

\newcommand{\quotes}[1]{``{#1}''}

\usepackage{fancyhdr}
\begin{document}

\newcolumntype{R}[1]{>{\raggedright\let\newline\\\arraybackslash\hspace{0pt}}m{#1}}
\newcolumntype{C}[1]{>{\centering\let\newline\\\arraybackslash\hspace{0pt}}m{#1}}
\newcolumntype{L}[1]{>{\raggedleft\let\newline\\\arraybackslash\hspace{0pt}}m{#1}}

\begin{frontmatter}
\title{Experimental data about the evacuation of preschool children from nursery schools - Part I: Pre-movement behaviour\footnote{This is a preprint version of the article accepted to Fire Safety Journal with DOI 10.1016/j.firesaf.2023.103798}}

\pagestyle{fancy}

\author[fsv,uceeb]{Hana Najmanov\'a}
\ead{hana.najmanova@cvut.cz}
\author[lund]{Enrico Ronchi}
\ead{enrico.ronchi@brand.lth.se}
\cortext[cor1]{Corresponding author. Tel. +420224357151, email: hana.najmanova@cvut.cz}
\address[fsv]{Czech Technical University in Prague, Faculty of Civil Engineering, Thakurova 7, 166 29 Prague, Czechia}
\address[uceeb]{Czech Technical University in Prague, University Centre for Energy Efficient Buildings, T\v rineck\'a 1024, 273 43 Bu\v st\v ehrad, Czechia}
\address[lund]{Lund University, Department of Fire Safety Engineering, John Ericssons v\"ag 1, 22363, Lund, Sweden}

\begin{abstract}
This article presents experimental data sets and information about the pre-movement behaviour and specific evacuation conditions observed in 15~evacuation drills in 10~nursery schools in the Czech Republic involving 970~children (3–-7~years of age) and 87 staff members. Based on the analysis of video recordings, over 1800~data points describing pre-movement times, the interpretation of warning signals, the reactions and activities of staff members and children, levels of physical assistance provided to the children, and exiting strategies were gathered, interpreted, and compared to key literature findings. Our observations showed that pre-movement times measured during the drills ranged from 3~to 59~s, depending on experimental conditions, and physical help was provided to 25\% of the participating children. The pre-movement behaviour of the participants was strongly influenced by the instructions provided to children as well as the daily routines, rules, and educational practices employed in the nursery schools. In addition, the principles of the affiliation model and role-rule model were identified as applicable to preschool children. To enable engineering applications (e.g., in evacuation modeling studies), findings are presented together with a set of nine behavioural statements regarding the pre-movement behaviour of preschool children.
\end{abstract}

 
\end{frontmatter}
\thispagestyle{fancy}

\paragraph{ Highlights }

\begin{itemize}
	\item New experimental data sets on pre-movement behaviour of children (3--7~years of age).
    \item 15~evacuation drills involving 970~children and 87~staff in 10~nursery schools.
    \item Data include pre-movement times and behaviour, assistance required, exiting strategies.
    \item Nine new behavioural statements for engineering applications are presented.
\end{itemize}

\newpage


\section{Introduction}
\label{sec:intro}
In evacuation design, the needs of population groups who may not be able to perform self-rescue activities must be carefully evaluated. Commonly referred to as \quotes{vulnerable} or \quotes{at risk} populations, such individuals may have various levels of self-rescue capability and may be dependent on the assistance provided by others \cite{kuligowski_human_2016, kuligowski_stair_2013,bukvic_review_2021}. Young children fall into this category, because their abilities for assessing emergency situations and self-rescue are limited  when compared to adults \cite{singer_piaget_1996}. During the early childhood period, children grow and progress at a rapid pace in all areas of development, including physical, motor, cognitive, social, and emotional \cite{berk_child_2006}. As they age, children progressively acquire  the ability to think, speak, learn, and reason. Attainments in cognition, emotions, and behaviours progress rapidly throughout the preschool years, and children, gaining new skills and experiences, grow in both intellectual and emotional ways. However, preschool children do not fully understand concrete logic and may be unable to manipulate information, which leads to limited thinking in very concrete terms \cite{singer_piaget_1996,shaffer_developmental_2010, schneider_memory_2015,siegler_childrens_1998}. In general, the levels of cognitive development in preschool children differ from those of adults and older children \cite{singer_piaget_1996}.   
Although a large corpus of research has been devoted to investigating the issue of fire evacuation for various occupancy, environmental, and emergency conditions (for example, \cite{predtechenskii_planning_1978,fruin_pedestrian_1971,pauls_review_1985, proulx_evacuation_1995,peacock_overall_2012}), little research to date has investigated the evacuation behavioural and movement characteristics of children under six years of age \cite{murozaki_study_1985,kholshevnikov_pre-school_2009,larusdottir_evacuation_2011,peacock_evacuation_2011-1,larusdottir_step_2011,campanella_empirical_2011,larusdottir_evacuation_2012,kholshevnikov_study_2012,capote_children_2012, cuesta_exploring_2013,takizawa_study_2013,taciuc_determining_2014,larusdottir_evacuation_2014,cuesta_collection_2015}. In recent years, this matter has attracted much attention from researchers in both fire safety science and pedestrian and evacuation dynamics \cite{,najmanova_experimental_2017,hamilton_human_2017,hamilton_toward_2019,fang_experimental_2019,li_comparative_2020, yao_research_2020, yao_childrens_2021, qi_social_2022}; nevertheless, the evacuation database in this field  remains limited \cite{hurley_engineering_2016}. \par
This article reports on a study that explored the evacuation process of children three--seven years of age accommodated in early childhood educational centres (referred to hereafter as \quotes{nursery schools}). The study focused on obtaining new experimental data sets and providing information regarding 15~evacuation drills (970~children and 87~staff members in ten participating nursery schools in the Czech Republic). Due to the large scope of research and the substantial amount of data collected, the findings of this study are presented in two separate publications that follow the general engineering timeline of fire evacuation from buildings~\cite{purser_behaviour_2003,ng_brief_2006}: the pre-movement phase (this paper) and movement phase (there is a paper part II associated with this work) of the evacuation process. This paper delves into the results regarding the pre-movement evacuation behaviour of children and staff members involved. The variables of interest include pre-movement times, interpretations of the warning signals, reactions and activities of staff members and children, levels of physical assistance provided to the children, and exiting strategies. For a look into the experimental findings describing children's movement characteristics and the movement behaviour arising from this study, the interested reader is referred to the associated paper part II. \par
Even though only a few research studies have surveyed the issue of pre-movement behaviour of preschool children  \cite{lovreglio_pre-evacuation_2019}, key findings regarding the behavioural aspects of preschool children during the pre-movement evacuation phase can be summarized as follows:
\begin{enumerate} [itemsep=1mm, parsep=0pt]
    \item Preschool children do not react independently to warning signals and pre-movement times are strongly influenced by the individual decisions and reactions of responsible staff members \cite{kholshevnikov_pre-school_2009, larusdottir_step_2011, kholshevnikov_study_2012, capote_children_2012, capote_children_2012-1, larusdottir_evacuation_2014, najmanova_experimental_2017, fang_experimental_2019}.
    \item The actions of staff members during pre-movement phase are dependent on the age of children \cite{larusdottir_evacuation_2014, najmanova_experimental_2017}. Therefore, longer \resub{pre-movement times} can be assumed for classes with younger children than for classes with older children \cite{najmanova_experimental_2017, hamilton_human_2017}.
    \item The evacuation behaviour of both children and staff members reflects the daily routines and rules in the different educational centres; children tend to follow a sequence of memorized daily activities (e.g., changing shoes, getting dressed, standing in front of the door waiting for the signal from staff members to go \cite{murozaki_study_1985, larusdottir_step_2011,larusdottir_evacuation_2014, najmanova_experimental_2017}).
    \item Preschool children tend to use familiar exits; however, the final exit choice is usually made by staff members \cite{larusdottir_step_2011,larusdottir_evacuation_2014, fang_experimental_2019}. 
  
\end{enumerate}

This study presents new experimental data sets, which enable appropriate future engineering applications (notably, evacuation modeling). A detailed background on the research methods and a comprehensive interpretation of the results is presented here together with a set of nine behavioural statements on the pre-movement behaviour of preschool children. Following the general concept of behavioural statements \cite{gwynne_guidance_2016,kuligowski_guidance_2017}, the presented behavioural statements aim to be of use when the pre-movement evacuation processes of preschool children need to be understood, quantified, and directly incorporated into life safety analyses (e.g., in evacuation modeling studies). The behavioural statements may also be used during the process of selecting evacuation scenarios and as a background for conducting sensitivity analysis when the assessment parameters are varied and their impacts are evaluated. For a better understanding of the evacuation processes for preschool children, this study investigates novel aspects describing exiting strategies and the types of supervision responsible adults provide to children during the pre-movement phase of the evacuation processes. Overall, this paper aims to promote an awareness of the specificity of the pre-movement evacuation of preschool children to fire engineers and provides data sets suitable for inclusion in fire safety and evacuation analyses. \par

\section{Research methods}
\label{sec:methods}
\resub{The data presented in this study were collected} during 15~evacuation drills carried out in 10~participating nursery schools in the Czech Republic (May and June~2019). To better analyze the behaviour of children in emergency scenarios, observing evacuation drills in nursery schools were selected as a suitable data collection method for our purposes \cite{haghani_crowd_2018, kinateder_virtual_2014}. Observation in the children's own environment was preferred to artificial or virtual setups in order to provide an acceptable balance between (semi-)naturalistic conditions, experimental control, repeatability, and the emotional and mental comfort of the participating children. \resub{In addition, large variability in participating nursery schools and children (see Section 2.1 for details) was considered to address the diversity of the population studied and to provide a global understanding of the evacuation of preschool children.}

\subsection{Participants and ethical considerations}\label{subsec:parti}
The 10 participating nursery schools varied in terms of their building design, capacity, experience with regular evacuation drills, and evacuation procedures. Hence, the boundary conditions of the evacuation drills differed considerably. An overview of the participating nursery schools is presented in Table~\ref{tab:schools}. To ensure the privacy of the participating subjects, any information which could enable identification of participants is not presented, and the nursery schools are labeled with letters. As seen from Table~\ref{tab:schools}, the overall capacity of the participating nursery schools varied from one class nursery schools to schools accommodating more than a hundred children; however, four class nursery schools were the most frequent type of schools in our study. \resub{Seven participating nursery schools (70\%) had some experience with an evacuation drill during the past five years: four nursery schools performed evacuation drills on a regular basis (annually or bi-annually), three nursery schools had only little drill experience (one evacuation drill in the last five years period) that was not positive, resulting in the decision not to conduct regular evacuation training anymore. Three participating nursery schools (30\%) had no prior experience with evacuation drills.} The participating nursery schools differed in types of employed warning signals used and their distribution in buildings. In Table~\ref{tab:schools}, \resub{\quotes{voice command} (VC)}; \quotes{manual sound signal} such as whistles or banging on gongs or pots \resub{(MSS); and \quotes{alarm} including siren and smoke detector signals (A)} are identified. \par

\begin{table}[!hbt]
    \footnotesize
    \centering
    \begin{tabular}{R{1cm}|R{2.5cm}|R{1.5cm}|R{2.2cm}|R{1.7cm}|R{1.1cm}|R{1.4cm}|R{1cm}}
      \hline
   
      \multirow{2}{=}{\textbf{School} \newline \textbf{label}}  & \multicolumn{4}{l|}{\textbf{Evacuation drills}} & \multicolumn{3}{l}{\textbf{Occupancy}}  \\
       \cline{2-8}
    
       &  \textbf{Prior experience} & \textbf{\resub{Serial No. of drill}} & \textbf{\resub{Evacuation type\textsuperscript{1)}}} & \textbf{Warning signal} & \textbf{Classes} & \textbf{Children} & \textbf{Staff}\\
       \hline
     A &   Yes (annually) & 1 & Announced & \resub{MSS} & 4 & 74 & 5 \\
     \hline
     B &  Yes (once in the last 5~years) & 1 & Unannounced & \resub{VC + MSS} & 11 & 207 & 21 \\
     \hline
     C  & Yes (bi-annually) & 1 & Semi-announced & \resub{VC + MSS} & 4 & 75 & 5 \\
     \hline
     \multirow{2}{=}{D} & \multirow{2}{=}{Yes (bi-annually)} & 1 & Announced & \resub{VC + MSS} & 4 & 63 & 4 \\
     \cline{3-8}
      &  &  2 & Semi-announced & \resub{VC + MSS} & 4 & 61 & 4 \\
     \hline
     E &  Yes (once in the last 5~years) & 1 & Unannounced & \resub{VC} & 6 & 100 & 9 \\
     \hline
     \multirow{2}{=}{F} & \multirow{2}{=}{No} & 1 & Announced & \resub{VC} & 1 & 20 & 2 \\
     \cline{3-8}
      &  & 2 & Announced & \resub{VC} & 1 & 23 & 2 \\
    \hline
     \multirow{2}{=}{G} & \multirow{2}{=}{Yes (once in the last 5~years)} & 1 & Announced & \resub{VC + MSS} & 3 & 52 & 4 \\
     \cline{3-8}
      &  & 2 & Semi-announced  & \resub{VC + MSS} & 3 & 50 & 3 \\
    \hline
     \multirow{2}{=}{H} & \multirow{2}{=}{No} & 1 & Announced & \resub{A} & 4 & 61 & 4 \\
     \cline{3-8}
      &  & 2 & Unannounced & \resub{VC + MSS} & 2 & 24 & 3 \\
    \hline
         \multirow{2}{=}{I} & \multirow{2}{=}{No} & 1 & Announced & \resub{A} & 1 & 12 & 3 \\
     \cline{3-8}
             &  & 2 & Semi-announced & \resub{A} & 1 & 12 & 3 \\
    \hline
    J &  Yes (annually) & 1 & Announced & \resub{A} & 8 & 136 & 16 \\
    \hline
    \multicolumn{5}{l}{\textbf{Total}} & \textbf{57} & \textbf{970} & \textbf{87}  \\
    \hline
    \multicolumn{8}{l}{\resub{\footnotesize \textsuperscript{1)} See Section 2.2 for details}}  \\
    \multicolumn{8}{l}{\resub{\footnotesize MSS = manual sound signal, VC = voice command, A = alarm}}  \\
    \hline
    \end{tabular}
    \caption{Overview of the participating nursery schools.}
    \label{tab:schools}
\end{table}
Children in the participating nursery schools attended both homogeneous and heterogeneous types of classes. For further analysis, different age groups were identified: Junior (3--4~years), Senior (5--6~years), Senior+ (6--7~years), and Mixed children (3--6~years). Due to ethical considerations, the ages of participating children were not registered individually; only an age range related to the whole class provided by the director of the participating nursery school was reported to researchers. An age-related overview of the participating children is given in Table~\ref{tab:age}.
\begin{table}[hbt!]
    \centering
    \footnotesize
    \begin{tabular}{R{3cm}|R{3cm}|R{3cm}|R{3cm}}
        \hline
   
       \textbf{Age group [years]} & \textbf{Label in this study} & \textbf{Number of classes} & \textbf{Number of children} \\
        \hline
        3--4 & Junior & 17 & 265 \\
        \hline
        5--6 & Senior & 15 & 295 \\
        \hline
        6--7 & Senior+ & 1 & 22 \\
        \hline
        3--6 & Mixed & 24 & 389 \\
        \hline
       \multicolumn{1}{l}{} & \multicolumn{1}{l}{\textbf{Total}} & \multicolumn{1}{l}{\textbf{57}} & \multicolumn{1}{l}{\textbf{970}} \\
       \hline
    \end{tabular}
    \caption{Age-related overview of the participating children.}
    \label{tab:age}
\end{table}
To ensure the fundamental ethical principles for research involving human subjects \cite{nilsson_exit_2009}, special attention was paid to appropriate ethical considerations as well as to the protection and respect given to the research participants.  \resub{The research project was approved by the Czech Technical University in Prague's ethical review board.} \resub{Informed consent was signed by nursery school directors in the planning phase of the study. In this phase, communication was also established with parents of children. Due to the large number of children involved, parental permission was obtained as opt-out consent, i.e., in a passive approval. Based on the information provided on the forthcoming evacuation drills and video recordings, parents had the opportunity to withdraw the participation in the study by contacting the school director or the authorised nursery school staff member. Staff members signed consent to use the data before (in the case of announced and semi-announced evacuation drills) or immediately after the experiment if the evacuation drill was not announced.}
\subsection{Experimental setup and equipment}\label{subsec:setup}
\begin{figure} [hbt]
    \centering
    \includegraphics[width=0.9\textwidth]{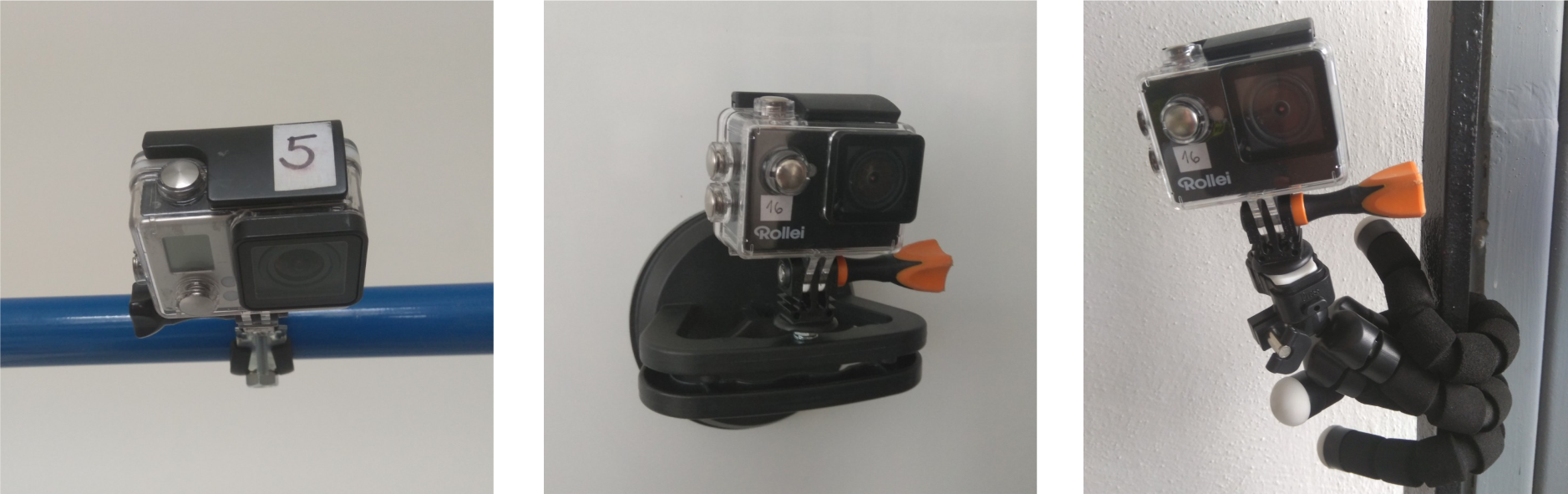}
    \caption{Example of mounting solutions for measurement equipment using telescopic building props in combination with pipe clamps (\textbf{left}), suction cups (\textbf{middle}), and flexible tripods (\textbf{right}).}
    \label{fig:camera}
\end{figure}

For each evacuation drill, a detailed experimental plan was compiled that included all relevant considerations for running the experiment (e.g., evacuation scenarios and procedures, population distributions, layouts of escape routes, and positioning and mounting of cameras). The study was entirely observational; the researchers did not intervene with the standard evacuation procedures for the nursery schools. 
The level of announcement and the amount of information provided to staff members and children was decided by the nursery school's director prior to the experiment. 
\resub{\quotes{The level of announcement} indicates to which extent the evacuation drills were announced to the participating staff members before it took place. The amount of information provided to children by parents and staff was not controlled; therefore, children could or could not be prepared for evacuation drills in advance. Only for unannounced drills parents were asked not to communicate the forthcoming events with their children.} Three different levels of announcement of drills took place in this study:
%
\begin{enumerate} [itemsep=1mm, parsep=0pt]
    \item \textbf{Announced evacuation drills:} Date and time of evacuation drills were available to all staff members prior to the drills.
    \item  \textbf{Semi-announced evacuation drills:} Evacuation drills were announced to all staff members beforehand; however, part of the information was not provided to them (e.g., date or/and time of the evacuation drill).
    \item \textbf{Unannounced evacuation drills:} No information about the evacuation drills was provided to staff members prior to the drills.
\end{enumerate}
Due to the diversity in information provided to participants, the impact of levels of preparedness of children and staff members could not be quantified in this study. 
All evacuation drills were recorded on video using outdoor digital cameras (resolution $720\times480$, frame rate 30~fps) temporarily installed in classrooms (only for announced and semi-announced evacuation drills) and on escape routes using different mounting equipment (e.g., suction cups, flexible tripods, multipurpose clamps; see Figure~\ref{fig:camera}). \resub{For unannounced evacuation drills, cameras were installed only in the corridors and all preparatory actions in the building were carried out fast and discreetly to avoid any contact between researchers and uninformed staff members. When necessary, the presence of strangers in the corridors was excused by an explanation decided by the school director.}
After finishing an evacuation drill, the acquired video footage was transferred to an encrypted storage device, synchronized, cut, labelled, and stored. Data were extracted manually using Avidemux 2.7.3 software that enabled a frame-by-frame analysis. Data were then organised in anonymous spreadsheets. 
\subsection{Limitations}\label{subsec:limit}
Interpretation of the data presented should be performed with the following limitations in mind: 
\begin{enumerate}[itemsep=1mm, parsep=0pt]
    \item \textbf{Naturalistic conditions}: The observation of evacuation drills as a data collection method paints only a simplified picture of real emergency situations and thus cannot reflect the full complexity of such situations in real life. Hence, the selected scenarios and procedures do not capture the diversity of real evacuation events (including factors such as availability of escape routes and the spread of fire and smoke) and their potential impacts on human movement and behaviour. 
    \item \textbf{Data collection and analysis methods}: The presented data sets are affected by uncertainties related to experimental measurements and statistical assumptions. Inaccuracies appear to originate from data extraction methods; the accuracy of the presented values for pre-movement times were conservatively limited to seconds. \resub{Due to the variability of the experimental conditions in the evacuation drills, in this study the impact of individual variables (such as different warning signals, evacuation procedures, staff-to-child ratios) on the pre-movement time is not investigated. In addition, quantification of time required by the participants for the various activities and instructions is not included in the pre-movement time analyses.}
    \item \textbf{Representativeness of data sets}: Due to the limited number of observations in certain age groups (e.g., only one class of Senior+ children), the data sets might not be fully representative \resub{for these groups}. 
\end{enumerate}

\section{Results} 
\label{sec:results}
The focus of investigation as reported in this paper was on evaluating the pre-movement times and evacuation behaviour of participants prior to exiting their initial place in the building (most often classrooms), not their movement through a building towards a place of safety outside. The latter behaviour and movement characteristics are articulated in the second paper related to this study. When cameras were installed in classrooms (i.e., for the announced and for some semi-announced evacuation drills), we also observed the decision-making and protection activities of staff members and children, the involvement of and instructions provided by staff members, and the way overall evacuation procedures were performed (e.g., the way children were organised). 
\subsection{Pre-movement times}\label{subsec:pre-mov}
Pre-movement times were measured in all participating nursery schools: 57~classes with 970~children and 87~staff members in total. Results are summarized in Table~\ref{tab:pre-mov}. Since the evacuation procedures during the drills were not controlled by the researchers and because the practices and fire documentation for participating nursery schools varied, the boundary conditions for the evacuation drills also differed considerably. Thus, Table~\ref{tab:pre-mov} provides background information including details of information provided to staff members and children prior to the experimental evacuation drills, warning signals, and placement of cameras in classrooms. \resub{\quotes{Pre-movement time} was defined as the time between the delivery of a warning signal (i.e., without warning interval) and the time when the first child in the class left the classroom door moving towards to the exit of the building. This definition reflects the specific evacuation conditions in nursery schools and considers the end of the pre-movement time as the time when children, already prepared for the next phase of the evacuation, received a clear signal to leave the classroom (that is, the children were considered as ready to leave by the responsible staff members). Hence, the pre-movement of the first child in a class was considered as representative for the entire class. This method for pre-movement time analysis that neglects individual response times is consistent with previous research focused on evacuation of preschool children \cite{kholshevnikov_study_2012, larusdottir_evacuation_2014}. To provide additional information on the time when classrooms were evacuated and to allow a comparison of the data sets presented with certain data from the literature, pre-movement times related to the last child in the class who left the classroom were also reported.}
\begin{landscape}
    \centering
    \footnotesize
\begin{longtable}{R{1.3cm}|R{3.5cm}|R{2.5cm}|R{3cm}|R{2,3cm}|R{1cm}|R{3,5cm}|R{3,5cm}}
    \caption{Pre-movement times and experimental background information.}
    \label{tab:pre-mov} \\  
    \hline
    \multirow[t]{2}{=}{\textbf{School and drill}} & \multicolumn{2}{R{6cm}|}{\textbf{Level of announcement: information provided to}}  & \multicolumn{2}{l|}{\textbf{Warning signal}} & \multirow[t]{2}{=}{\textbf{IN}} & \multicolumn{2}{R{7cm}}{\textbf{Pre-movement time \newline (mean/min/max/SD) [s] (data points)}}  \\
    \cline{2-5} \cline{7-8}
        \endfirsthead
     \hline \endhead
     \hline \endfoot
     \hline \endlastfoot
     & \textbf{Staff} & \textbf{Children} & \textbf{Signal} & \textbf{Method} & & \textbf{First child} & \textbf{Last child} \\
    \hline
    A1 & A: Date and approx. time & Motivated by a game beforehand\textsuperscript{1)} & Gong & P in corridors & Yes & 28 / 16 / 40 / 10 (4) &  25 / 60 / 15 (4) \\
    \hline
    B1 & U: approx. month & No info & Verbal, afterwards verbal and whistle & Broadcast, then P in corridors. &  No & 33 / 15 / 59 / 14 (11) & 45 / 27 / 77 / 16 (11) \\
    \hline
    C1 & SA: Date & No info\textsuperscript{1)} & Pot banging & P in corridors & Yes & 9 / 4 / 11 /3 (4) & 20 / 12 / 25 / 5 (4)\\
    \hline
    D1 & A: Date and approx. time & No info\textsuperscript{1)} & Verbal and whistle & P in corridors & Yes & 29 / 17 / 46 / 14 (4) & 45 / 22 / 65 / 23 (4) \\
    \hline
    D2 & SA: Date & No info\textsuperscript{1)} & Verbal and whistle & P in corridors & Yes & 26 / 10 / 41 / 13 (4) & 42 / 26 / 56 / 13 (4) \\
    \hline
    E1 & U: approx. month & No info & Verbal & Broadcast & No & 38 / 25 / 54 / 10 (6) &  49 / 37 / 62 / 11 (6) \\
    \hline
    F1 & A: Date and time & No info & Verbal & P in classroom & Yes & 19 / - / - / -  (1) & 29 / - / - / -  (1) \\
    \hline
    F2 & A: Date and time & No info & Verbal & P in classroom & Yes  & 5 / - / - / - (1) & 26 / - / - / -  (1)  \\
    \hline
    G1 & A: Date and approx. time & No info & Verbal and triangle & P in corridors & No\textsuperscript{2)} & 5 / 3 / 6 / 2 (3) & 21 / 17 / 28 / 6 (3)  \\
    \hline
    G2 & SA: Date & No info & Verbal and triangle & P in corridors & No\textsuperscript{2)} & 9 / 7 / 13 / 3 (3) & 20 / 16 / 24 / 4 (3)\\
    \hline
    H1 & A: Date and approx. time & No info & Verbal and megaphone's siren &  P in corridors & No & 14 / 8 / 19 / 5 (4) & 21 / 14 / 29 / 7 (4) \\
    \hline
    H2 & U: No info & No info & Verbal and pot banging & P in corridors & No & 28 / 10 / 45 / 25 (2) & 41 / 27 / 55 / 20 / (2) \\
    \hline
    I1 & A:Date and approx. time & No info & Smoke detection signal & P in corridors & Yes & 13 / - / - / -  (1) & 19 / - / - / -  (1)   \\
    \hline
    I2 & SA: Date & No info & Smoke detection signal & P in corridors & Yes &  11 / - / - / - (1) & 16 / - / - / -  (1)  \\
    \hline
    J1 & A: Date and time & Date and time & Megaphone's siren & P in corridors & Yes & 22 / 8 / 46 / 14 (8) & 32 / 17 / 61 / 15 (8) \\
    \hline
    \multicolumn{6}{l}{\textbf{Total}} & \multicolumn{1}{l}{\textbf{23 / 3 / 59 / 15 (57)}} &  \multicolumn{1}{l}{\textbf{35 / 12 / 77 / 16 (57)}}\\
    \hline
    \multicolumn{8}{l}{\footnotesize A: announced, SA: semi-announced, U: unannounced} \\
    \multicolumn{8}{l}{\footnotesize{P: Personally by the responsible staff member; IN: Cameras placed inside the classrooms}}\\
    \multicolumn{8}{l}{\footnotesize \textsuperscript{1)} Exceptions could occurred in particular classes in the nursery school based on different approach taken by staff members} \\
    \multicolumn{8}{l}{\footnotesize \textsuperscript{2)} Cameras not placed in classrooms, observations in classrooms enabled thanks to glass walls} \\
    \hline
\end{longtable}
\end{landscape}
\subsubsection{Age of children}\label{subsub:age}
The pre-movement times measured in the study organized according to the different participating age groups (Junior, Senior, Senior+, and Mixed) are summarized \resub{in Table~\ref{tab:pre-mov_age}}. Based on the results, no apparent correlation between pre-movement time and the age of children could be determined. \resub{This can be attributed to both the limited number of data points collected (each pre-movement time represents the pre-movement phase for an entire group of children in the classroom) and the variability of conditions in the evacuation drills that limited a deeper analysis of single variables that affected the pre-movement time.} \par 
\begin{table}[hbt]
    \centering
    \footnotesize
    \begin{tabular}{R{2.5cm}|R{4cm}|R{4cm}}
    \hline
    \multirow{2}{=}{\textbf{Age group}} & \multicolumn{2}{R{8cm}}{\textbf{Pre-movement time \newline (mean/min/max/SD) [s] (data points)}} \\
    \cline{2-3}
    & \textbf{First child} & \textbf{Last child} \\
    \hline
    Junior &  30 / 10 / 59 / 16 (17) & 41 / 15 / 77 / 17 (17) \\
    \hline
    Senior & 20 / 4 / 54 / 15 (15) & 32 / 12 / 62 / 17 (15) \\
    \hline
    Senior+ & 34 / - / - / (1) & 44 / - / - / - (1) \\
    \hline
    Mixed & 20 / 3 / 46 / 13 (24) & 33 / 16 / 65 / 16 (24) \\
    \hline
    \end{tabular}
       \caption{Pre-movement times in different age groups.}
       \label{tab:pre-mov_age}
\end{table}
\subsubsection{Level of announcement and participant preparedness}\label{}
\begin{figure} [hbt!]
    \centering
    \includegraphics[width=0.6\textwidth]{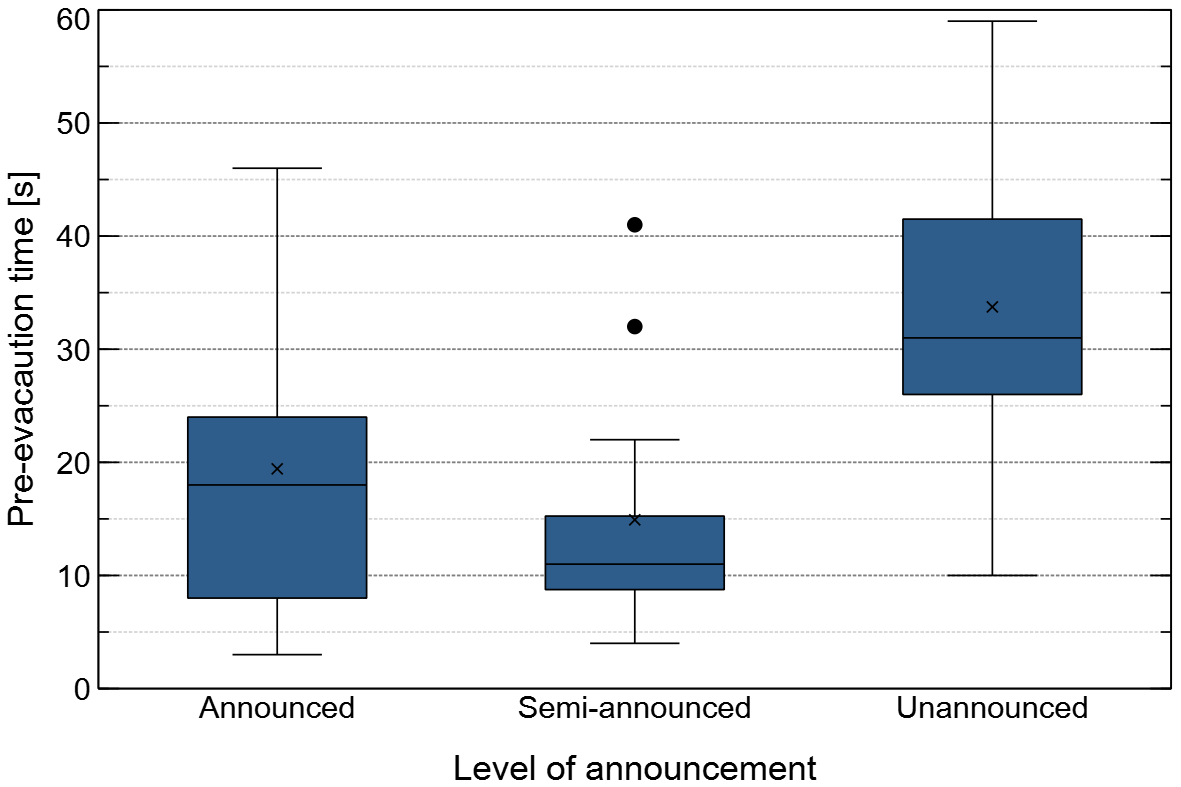}
    \caption{Box plots describing pre-movement times in relation to different levels of announcement for the experimental evacuation drills.}
    \label{fig:pre-mov_ann}
\end{figure}

Different levels of information were provided to staff members and children prior the evacuation drills in the participating nursery schools, as noted above. Besides, the approach taken by staff members to prepare children for the drills could vary amongst classes even in the same nursery school. Thus, children exhibited different levels of preparedness for the evacuation drills based on the decisions of staff members in this regard. Pre-movement times observed in relation to different levels of announcement and preparedness of children are summarized \resub{in Table~\ref{tab:pre-mov_ann}} \resub{and is shown using box plots in Figure~\ref{fig:pre-mov_ann}. Longer pre-movement times were observed in the unannounced evacuation drills when the participants were not aware of the drills in advance. The significant difference between the medians of the pre-movement times observed in different evacuation types was confirmed using the Kruskal-Wallis Test (H$=$15.23, p$=0.001$) and the Mood’s Median Test (p$=0.004$). The null hypothesis that the medians of the group population are equal was rejected at the 0.05 significance level. Furthermore, shorter pre-movement times appeared to occur in evacuation drills when children were prepared by staff members for the drills in advance. However, this finding was not statistically proven due to the limited number of data points in the observed groups.}. 
\par
\begin{table}[hbt]
    \centering
    \footnotesize
    \begin{tabular}{R{3cm}|R{3.7cm}|R{3.5cm}|R{3.5cm}}
    \hline
    \multirow{2}{=}{\textbf{Level of announcement}} & \multirow{2}{=}{\textbf{Preparedness of children}} & \multicolumn{2}{C{7cm}}{\textbf{Pre-movement time \newline (mean/min/max/SD) [s] (data points)}} \\
    \cline{3-4}
    & & \textbf{First child} & \textbf{Last child} \\
    \hline
    \multirow{2}{=}{Announced} & Children prepared & 13 / 8 / 24 / 7 (8) & 22 / 14 / 37 / 8 (8) \\
    \cline{2-4}
    & Children not prepared & 22 / 3 / 46 / 14 (18) & 36 / 17 / 65 / 16 (18) \\ 
    \hline
    \multirow{2}{=}{Semi-announced} & Children prepared & 7 / 4 / 9 / 4 (2) & 16 / 12 / 19 / 5 (2)  \\
    \cline{2-4}
    & Children not prepared & 17 / 7 / 41 / 11 (10) & 29 / 16 / 56 / 13 (10) \\ 
    \hline
    Unannounced & Children not prepared & 34 / 10 / 59 / 13 (19) & 46 / 27 / 77 / 15 (19) \\
    \hline
    \end{tabular}
        \caption{Pre-movement times in relation to level of announcement and preparedness of children.}
            \label{tab:pre-mov_ann}
\end{table}
\subsection{Interpretation of warning signal}\label{subsec:interpreataion}
During the announced and semi-announced evacuation drills, staff members reacted promptly and adequately, showing no difficulty interpreting the warning signals. In contrast, for unannounced evacuation drills, the warning signals were ignored several times by staff members because the broadcasts were difficult to hear or staff members did not know what the warning signals meant. This led to considerably longer pre-movement times (up to 35~s delay times). \par
\subsection{Reaction and activities of the participants}\label{subsec:reaction}
After interpreting warning signals, it appears from analysis that staff members reacted quickly for the most part, beginning to instruct the children about where to go and how to behave. The most common activities and instructions given by staff members to children are summarized separately for the different age groups in Table~\ref{tab:activities}.\par
Children's reactions and activities were analyzed for 39~classes. 90\% of 649~observed children responded to instructions obtained from the staff members quickly and without any complications. Only 10\% of children reacted slowly; ignored the instructions provided; required repetition of advice, or needed additional physical help such as to be pushed, taken by the hand, lifted from a table, or carried. It should be noted that such behaviour was observed only in Junior children \resub{(3.1\% of all Junior children)} and in younger children present in the Mixed classes \resub{(15.5\% of all Mixed children)}. These children typically continued to play with toys, stayed in place, or struggled to pick up their toys. In a few cases, these children received physical help from older children in the same group. Moreover, several children were seen following usual activities from their daily routines, notably putting away their toys, exercise aids, or cushions away or even going back to the classroom to return things that had remained in their hands during the evacuation process.  \par
\begin{table}[hbt!]
    \centering
    \footnotesize
    \begin{tabular}{R{5cm}|R{1,5cm}|R{1,5cm}|R{1,5cm}|R{1,5cm}|R{1,5cm}}
    \hline
   
    \multirow{2}{=}{\textbf{Activities}} & \multicolumn{5}{R{7,5cm}}{\textbf{Frequency [\%] (data points)}} \\
    \cline{2-6}
  
    & \textbf{Total} & \textbf{Junior} & \textbf{Senior} & \textbf{Senior+} & \textbf{Mixed} \\
    \hline
    Go to the exit and open the door & 55.0 (40) & 50.0 (10) & 12.5 (8) & N/A & 27.3 (22) \\
    \hline
    Take a roll call & 45.6 (57) & 41.2 (17) & 73.3 (15) & 0.0 (1) & 33.3 (24) \\
    \hline
    Explain the situation briefly & 40.0 (40) & 50.0 (10) & 12.5 (8) & N/A & 45.5 (22) \\
    \hline
    Count children & 32.6 (46) & 44.4 (9) & 16.7 (12) & 0.0 (1) & 37.5 (24) \\
    \hline
    Take keys/mobile phone & 27.6 (57) & 23.5 (17) & 26.7 (15) & 0.0 (1) & 25.0 (24) \\
    \hline
    Turn off lights, close the door & 7.8 (51) & 7.1 (14) & 8.3 (12) & 0.0 (1) & 8.3 (24)  \\
    \multicolumn{3}{c}{} \\
 
    \multirow{2}{=}{\textbf{Instructions}} & \multicolumn{5}{R{7,5cm}}{\textbf{Frequency [\%] (data points)}} \\
     \cline{2-6}
    
    & \textbf{Total} & \textbf{Junior} & \textbf{Senior} & \textbf{Senior+} & \textbf{Mixed} \\
    \hline
    Instruct children to be quick  & 57.5 (40) & 50.0 (10) & 37.5 (8) & N/A & 68.2 (22) \\
    \hline
    Instruct children to form pairs and hold each other's hands & 38.6 (57) & 58.8 (17) & 46.7 (15) & 100.0 (1) & 16.7 (24) \\
    \hline
    Instruct children which exit to use & 30.0 (40) & 50.0 (10) & 12.5 (8) & N/A & 27.3 (22) \\
    \hline
    Instruct children to make a formation & 27.5 (40) & 40.0 (10) & 25.0 (8) & N/A & 22.7 (22) \\
    \hline
    Instruct children to wait at a specific place & 12.5 (40) & 0.0 (10) & 12.5 (8) & N/A & 18.2 (22) \\
    \hline
    Instruct children to go slowly & 10.0 (40) & 0.0 (10) & 37.5 (8) & N/A & 4.5 (22) \\
    \hline
    Instruct children be quiet/not to scream/cry & 10.0 (40) & 30.0 (10) & 0.0 (8) & N/A & 4.5 (22) \\
    \hline
    Instruct children to leave toys & 10.0 (40) & 0.0 (10) & 0.0 (8) & N/A & 18.2 (22) \\
    \hline
    \end{tabular}
    \caption{Overview of the instructions and activities performed by the staff members in different age groups during the evacuation drills. }
    \label{tab:activities}
\end{table}
\subsection{Level of physical assistance provided to children} \label{subsec:assistance}
Physical support activities provided by the staff members to children were analyzed during the pre-movement phase for 40~classes. Additional observation of physical contact between the staff members and children was possible using recordings from cameras placed in the corridors when the door of the classroom was opened. To describe the differences in the type of help received, five categories were established, see Table~\ref{tab:help}. The category \resub{\quotes{No physical assistance}} was the most frequent (75\%) and includes all cases where \resub{no physical contact} between a child and a staff member was noticed. The most common kind of physical contact was \quotes{Gentle pushing} on children's backs, shoulders, arms, or heads, which was observed in 18.7\%. However, this contact was not always necessary to proceed the evacuation process, i.e., contact was also performed by staff members in the cases where children reacted and behaved orderly, and independently not requiring any assistance. This type of contact was mostly observed when a roll call of children was taken or during exiting through a classroom door. Physical contact necessary for performing a successful evacuation (\quotes{Physical contact needed}, e.g., redirecting and pushing a child to the right exit, picking a child up from a table) was observed for 3.6\% of children. A small percentage of the children was taken by the hand (\quotes{Hand holding}, 2.3\%) and only several children (0.3\%) had to be carried during the pre-movement phase (\quotes{Carried)}.   \par    
\begin{table}[hbt!]
    \centering
    \footnotesize
    \begin{tabular}{R{4cm}|R{1,7cm}|R{1,7cm}|R{1,7cm}|R{1,7cm}|R{1,7cm}}
 
    \hline
    \multirow{2}{=}{\textbf{Physical assistance provided}} & \multicolumn{5}{R{10.2cm}}{\textbf{Frequency [\%] (data points)}} \\
    \cline{2-6}
  
    & \textbf{Total} & \textbf{Junior} & \textbf{Senior} & \textbf{Senior+} & \textbf{Mixed} \\
    \hline
    \resub{No physical assistance} & 75.0 (441) & 73.7 (81) & 89.2 (121) & 81.8 (18) & 67.7 (244) \\
    \hline
    Gentle pushing & 18.7 (181) & 20.8 (55) & 10.2 (30) & 18.2 (4) & 23.7 (92)   \\
    \hline
    Physical contact needed & 3.6 (24) & 2.1 (3) & 0.0 (0) & N/A & 5.7 (21) \\
    \hline
    Hand holding & 2.3 (15) & 2.8 (4) & 0.7 (1) & N/A & 2.7 (10) \\
    \hline
    Carried &  0.3 (2) & 0.7 (1) & 0.0 (0) & N/A & 0.3 (1) \\
    \hline
    \end{tabular}
    \caption{Level of physical assistance provided to children in different age groups during the evacuation drills.}
    \label{tab:help}
\end{table}
\subsection{Exiting strategies}\label{subsec:leaving}
\begin{figure} [hbt!]
    \centering
    \includegraphics[width=0.8\textwidth]{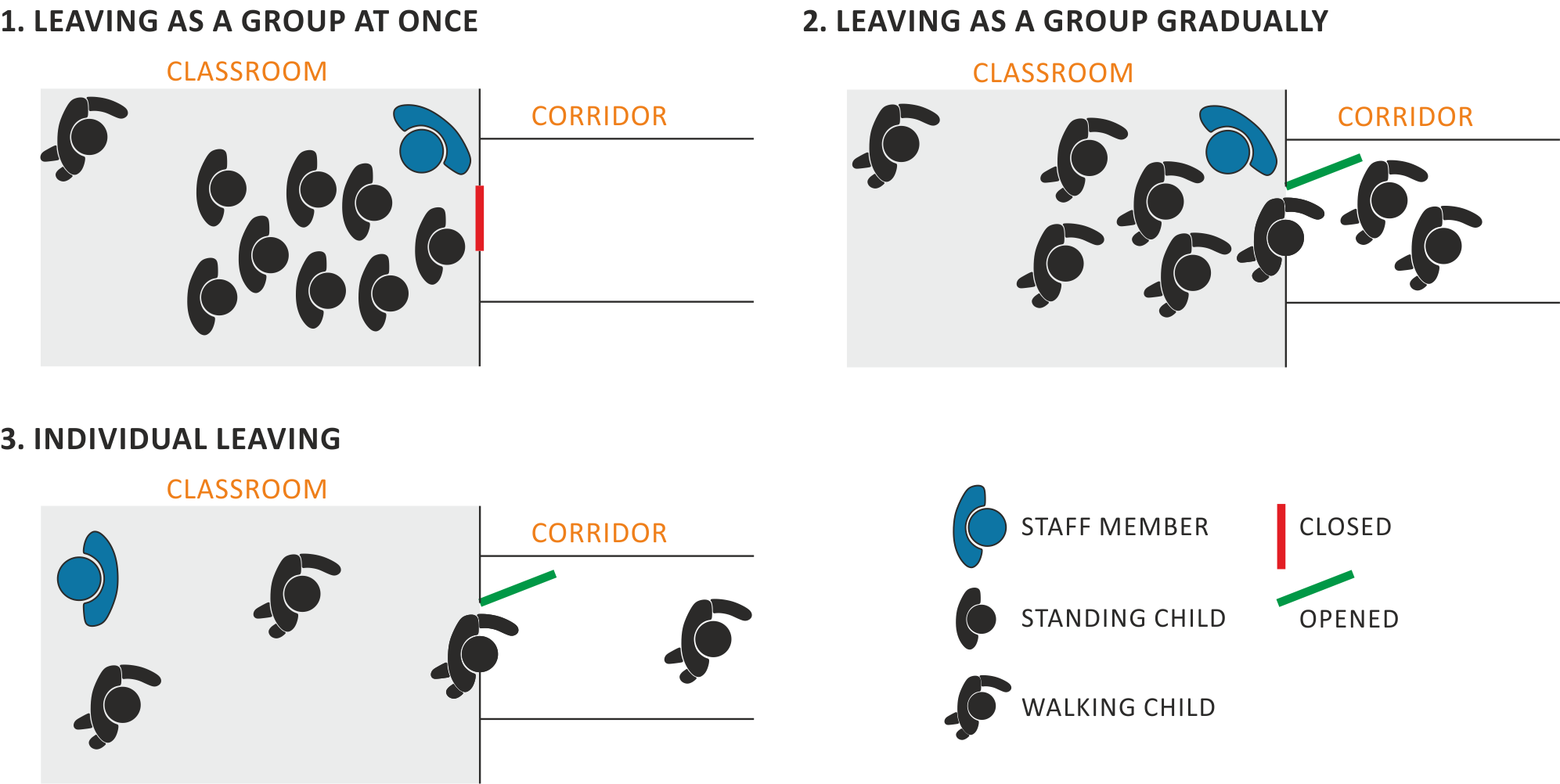}
    \caption{Exiting strategies observed during the evacuation drills.}
    \label{fig:leaving}
\end{figure}

\begin{figure} [hbt!]
    \centering
    \includegraphics[width=0.85\textwidth]{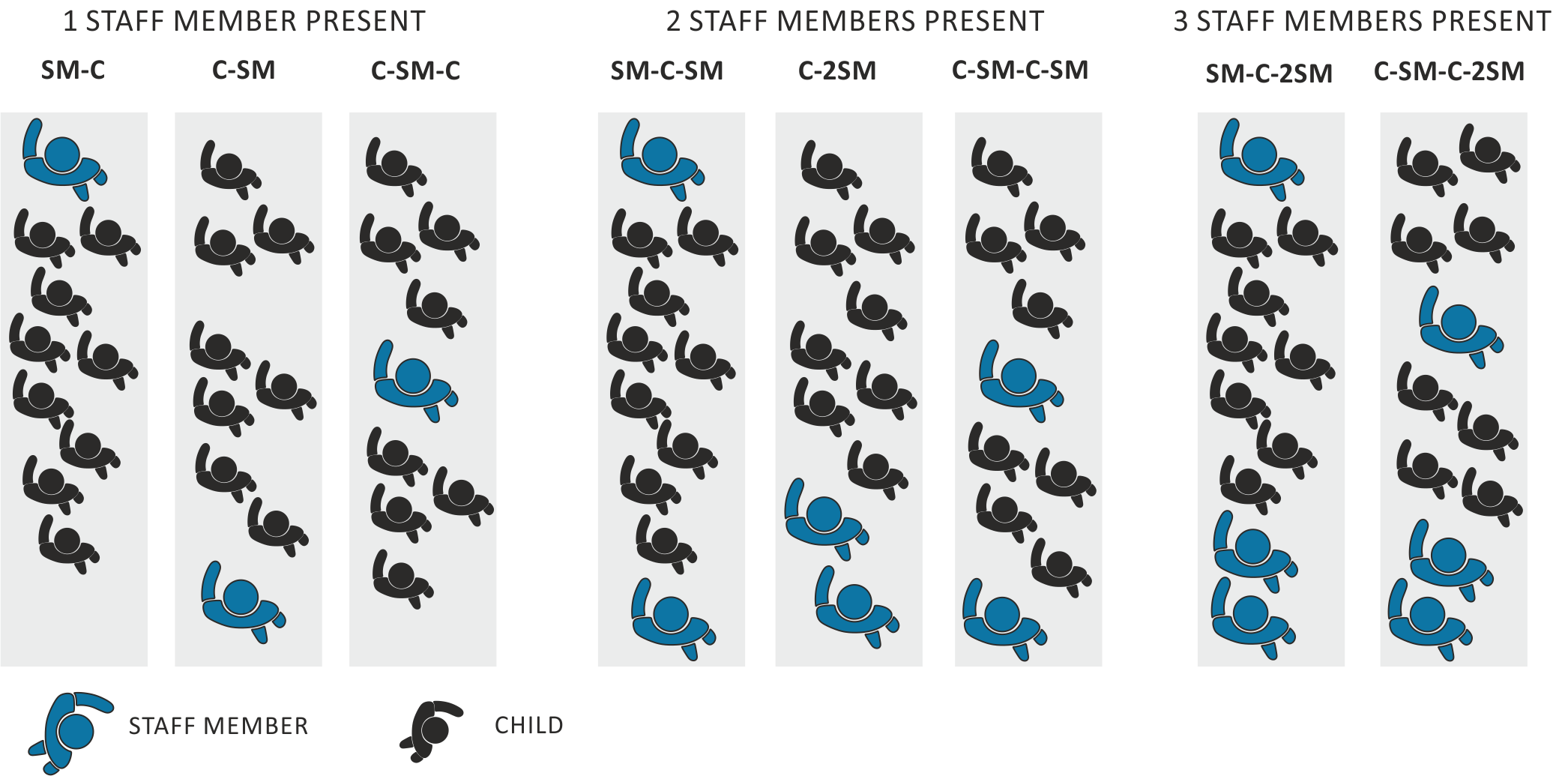}
    \caption{Various positions of staff members in the exiting group in relation to number of staff members present in a class.}
    \label{fig:position}
\end{figure}

Based on the observations, three organizational strategies staff members took while preparing children for evacuation and when exiting the classroom were identified (see Figure~\ref{fig:leaving} for a graphical interpretation):   
\begin{enumerate} [itemsep=1mm, parsep=0pt]
    \item \textbf{Exiting as a group at once:} All children were gathered in front of a closed exit from the classroom and formed a standing queue. Once the whole group was accounted, the door was opened and children started to leave the classroom. 
    \item \textbf{Exiting gradually as a group:} All children left the classroom together; however, they were not gathered and accounted in front of a classroom exit. When children arrived at the door as a moving queue, the door was already open and staff members supervised the process so that the children leave the room in an organized fashion.
    \item \textbf{Individual exiting:} Supervised from a distance, children left the classroom individually and were instructed to wait at a specified place inside or outside the building. 
\end{enumerate}
In 58\% of the evacuation drills, children left the classroom as a compact group; in 19\% of the drills they left as a supervised group, gradually (without waiting until the group was complete); and in 23\% of the drills, children were allowed to exit the classroom individually. The latter situations were observed most often for one class nursery schools or for nursery schools with a simple building layout and with a short evacuation route ending outdoors.
Another aspect related to the exiting strategies was the position of a staff member in the exiting group, i.e., if the first (or last) person leaving the classroom who led (followed up) the group was a staff member or a child. Observations of all 57~participating classes showed that in 61.4\% the first person leaving a classroom was a child and in 38.6\% a staff member. By contrast, a child as the last person to leave a classroom was observed only in 21.1\% of classes in this study, whereas a staff member exited last in the remaining 78.9\% of classes. It can be assumed that the position of a staff member in an exiting group was to a large extent dependent on the number of staff members present in a class. Regarding this, different exiting strategies and their frequencies during evacuation drills are listed in Table~\ref{tab:position} and illustrated in Figure~\ref{fig:position}.
\begin{table}[hbt!]
    \centering
    \footnotesize
    \begin{tabular}{R{1,5cm}|R{1,8cm}|R{2,1cm}|R{6cm}|R{1,8cm}}
    \hline

     \textbf{Number of staff members} &\textbf{Frequency [\%] (data points)}  &  \textbf{Symbolic description} & \textbf{Explanation} & \textbf{Frequency [\%] (data points)} \\
      \hline
      \multirow{3}{=}{1} & \multirow{3}{=}{59.7 (33)} & SM-C & Staff member left the classroom as the first person followed by children & 27.3 (9) \\
      \cline{3-5}
     &  & C-SM & Staff member left the classroom as the last person following all children & 63.6 (21) \\ 
      \cline{3-5}
     & & C-SM-C & Staff member left the classroom following several children before the remaining children & 9.1 (3) \\
     \hline
      \multirow{3}{=}{2} & \multirow{3}{=}{31.6 (18)} & SM-C-SM & Staff members left the classroom as the first and last person closing the group of children from both sides & 61.1 (11) \\
      \cline{3-5}
      & & C-2SM & Both staff members left the classrooms as two last persons following all children & 22.2 (4) \\
      \cline{3-5}
      & & C-SM-C-SM & The first staff member left the classroom following several children, the another staff member left the classroom as the last person & 16.7 (3)  \\
      \hline
      \multirow{2}{=}{3} & \multirow{2}{=}{10.5 (6)} & SM-C-2SM & Staff members left the classroom as the first person and as two last persons closing the group of children from both sides & 33.3 (2) \\
      \cline{3-5}
      & & C-SM-C-2SM & The first staff member left the classroom following several children, other two staff members left the classroom as two last persons & 66.7 (4)  \\
      \hline
      \multicolumn{5}{l}{\footnotesize{SM: staff member}} \\
      \multicolumn{5}{l}{\footnotesize{C: children}} \\
      \hline
    \end{tabular}
    \caption{Position of a staff member in the leaving group in relation to the number of the staff members present in a class.}
    \label{tab:position}
\end{table}
When the class was supervised by only one staff member, staff members had to decide whether to lead or to follow behind the children's flow. In the \quotes{SM-C} case (see Table~\ref{tab:position} and Figure~\ref{fig:position}), the staff member could control a proper selection of escape routes and took the role of the leader. However, the children in the classroom and the tail end of the children's group remained unsupervised and the staff member began losing sight of all children, making it difficult to determine whether someone was missing. These disadvantages were eliminated in strategies where a staff member decided to leave the classroom last (\quotes{C-SM}, see Table~\ref{tab:position} and Figure~\ref{fig:position}). This strategy proved to be effective in classrooms with a direct exit to an outdoor space or where a single evacuation route was available. Nevertheless, observations showed that if children were leading a whole group, they could easily take the wrong direction and used undesirable evacuation routes. Consequently, an ideal strategy appears to be having enough staff members to both lead and follow behind the flow children. Remarkably, this option was not always observed during analysis of the evacuation drills. Furthermore, it was not unusual for the first staff member to leave the classroom somewhere in the middle of the group (e.g., \quotes{C-SM-C} in Table~\ref{tab:position} and Figure~\ref{fig:position}). This situation was typically caused by a self-closing door that had to be held open by the staff member initially positioned at the head of the flow of children. 

\section{Discussion}\label{sec:discuss}
In this section, the presented results and findings are discussed in the context of existing research focused on children’s pre-movement behaviour. A new set of behavioural statements on the evacuation behaviour of preschool children in early childhood education centres is introduced and discussed. 

\subsection{Comparison with previous research studies}\label{subsec:comparison}
Human behaviour and response activities during the pre-movement phase of an evacuation process are often quantified using the variable of pre-movement time. A comparison of the current data set describing pre-movement times with data available in the literature \cite{larusdottir_evacuation_2014, kholshevnikov_pre-school_2009, capote_children_2012, capote_children_2012-1, najmanova_experimental_2017, hamilton_human_2017} is summarized in Table~\ref{tab:pre-mov_comp} including the information regarding the level of announcement given to the participants and exiting strategies. 
\begin{table}[hbt!]
    \centering
    \footnotesize
    \begin{tabular}{R{1.5cm}|R{2.7cm}|R{2.2cm}|R{0.8cm}|R{2.8cm}|R{3.6cm}}
    \hline
    \multirow{2}{=}{\textbf{Reference}} & \multirow{2}{=}{\textbf{Announcement (number of drills)}} & \multirow{2}{=}{\textbf{Exiting strategy}} & \multirow{2}{=}{\textbf{Age [years]}} &\multicolumn{2}{R{6cm}}{\textbf{Pre-movement time \newline (mean/min/max/SD) [s] (data points)}} \\ 
    \cline{5-6}
    & & & & \textbf{First child} & \textbf{Last child} \\
    \hline
    Larusdottir et al.~\cite{larusdottir_evacuation_2014} &  Partly announced (16) & N/A & 3--6 & N/A & 114 / 10 / 545 / - (N/A) \newline 54 / - / - / - (N/A) \newline (excluding warning time)  \\
    \hline
    Kholshchevnikov et al.~\cite{kholshevnikov_pre-school_2009} & N/A  & N/A \newline (indoor clothing)  & 3--7 & N/A  &  8 / 4 / 18 / 3 (N/A) \newline (indoor clothing) \newline 23 / 8 / 48/ 12 (N/A) \newline (with blankets) \newline 196 / 126 / 270 / 47 (N/A) \newline (winter clothing) \\
    \hline
    \multirow{2}{=}{Capote et al.~\cite{capote_children_2012, capote_children_2012-1}} & \multirow{2}{=}{Announced to staff members (1)} & \multirow{2}{=}{Group at once} & 4--5 & 38 / - / - / - (1 class) & \multirow{2}{=}{N/A}  \\
    \cline{4-5}
    & & & 5--6 & 21 / - / - / - (1~class) \\
    \hline
    \multirow{2}{=}{Najmanová and Ronchi \cite{najmanova_experimental_2017}} & \multirow{2}{=}{Semi-announced (2)} & \multirow{2}{=}{Group at once} & 3--4 & 46 / 12 / 68 / 16 (6~classes) & N/A  \\
    \cline{4-6}
    & & & 5--6 & 20 / 13 / 32 / 7 (4~classes) & \\
    \hline
    Hamilton et al.~\cite{hamilton_human_2017} & Announced (1) and unannounced (3) & N/A  & 5--6 & 23\textsuperscript{1)}/5\textsuperscript{1)}/46\textsuperscript{1)}/- (24~classes) & N/A  \\
    \hline
    \multirow{3}{=}{Current study} & Announced (8) \newline Semi-announced (4) \newline Unannounced (3) & -- & \multirow{3}{=}{3--7} & 19 (26~classes) \newline 15 (12~classes) \newline 34 (19~classes) & 31 (26~classes) \newline 27 (12~classes) \newline 46 (19~classes) \\
    \cline{2-6}
    & -- & Group at once \newline Group gradually \newline Individually &\multirow{3}{=}{3--7} &  31 (33~classes) \newline 14 (11~classes) \newline 11 (13~classes) & 43 (33~classes) \newline 24 (11~classes) \newline 24 (13~classes) \\
    \cline{2-6}
    & \multicolumn{3}{l|}{In total (15)} & 22 & 34 \\
    \hline
    \multicolumn{6}{l}{\footnotesize{N/A information not available}}\\
    \multicolumn{6}{l}{\footnotesize \textsuperscript{1)} the value read from the graph}\\
    \hline
    \end{tabular}
    \caption{Comparison of the observed pre-movement times with available literature data.}
    \label{tab:pre-mov_comp}
\end{table}
In the introduction to this paper, several behavioural aspects related to preschool children's evacuation behaviour during the pre-movement phase were highlighted (see Section~\ref{sec:intro}). A summary of the presented outcomes in relation to those findings is provided in Table~\ref{tab:behaviour} and discussed briefly below. \par
\begin{table}[hbt!]
    \centering
    \footnotesize
    \begin{tabular}{R{9cm}|R{5cm}}
      \hline
      \textbf{Finding [references]}   &  \textbf{The current study }\\
      \hline
      Preschool children do not react to a warning signal independently \cite{kholshevnikov_pre-school_2009, larusdottir_step_2011, kholshevnikov_study_2012, capote_children_2012, capote_children_2012-1, larusdottir_evacuation_2014, najmanova_experimental_2017, fang_experimental_2019} & Confirmed with exceptions \\
      \hline
      Actions of staff members during pre-movement phase are dependent on age of children \cite{larusdottir_evacuation_2014, najmanova_experimental_2017} & Confirmed \\
      \hline
      Longer \resub{pre-movement times} can be assumed in classes with younger children than in classes with older children \cite{najmanova_experimental_2017, hamilton_human_2017} & Neither confirmed nor rebutted \\
      \hline
      Evacuation behaviour reflects the daily routine and rules; preschool children tend to follow a sequence of memorized as part of daily activities \cite{murozaki_study_1985, larusdottir_step_2011,larusdottir_evacuation_2014, najmanova_experimental_2017} & Confirmed \\
      \hline
      Preschool children tend to use familiar exits; the final exit choice is performed by staff members \cite{larusdottir_step_2011,larusdottir_evacuation_2014, fang_experimental_2019} & Confirmed \\
           \hline
    \end{tabular}
           \caption{Summary of behavioural aspects considering the evacuation behaviour of preschool children during the pre-movement phase presented in this study and in the literature.}

    \label{tab:behaviour}
\end{table}
The hypothesis that preschool children react first to instructions given by adults and not to the warning signals as such was confirmed in most of the nursery schools and classes observed in this study. Children in two classes at only one nursery school started to evacuate individually following the voice message in the corridors provided by the director who instructed the children to leave the building. The children's reaction in that case may be explained as a result of daily practices at that specific nursery school since children at that school are used to moving through the building without supervision at any time. It appears from such observations that common school practices as well as educational programs and overall approaches to children in early childhood education centres may considerably influence reactions and behaviour in emergency situations. \par   
The observations of evacuation behaviour during the pre-movement phase further revealed that the activities taken by staff members and their instructions given to the children varied with the age of children. In Junior classes and in Mixed classes where children between of three --four~years of age were accommodated, staff members often instructed the children about which exit to use, explaining the situation to them and counting them before leaving the classroom. Moreover, younger children were often asked to be quiet or to stop playing and to follow given instructions having to be repeated many times. The assumption from the available literature that young children require more care and attention from staff members was fully confirmed in this study. Slow reaction to the instructions provided was observed only for younger children, who required also additional physical help. The evacuation behaviour of staff members followed a concept known as the role-rule model \cite{canter_domestic_1980}. According to this model, the behaviour of individuals during emergencies is guided by their \quotes{role in daily life} (i.e., a set of expectations) in which they expect themselves to be in the actual context, i.e., the coping strategies and engaged activities undertaken by staff members were determined by their social roles and the positions of the responsible adults. \par 
\resub{Conversely, due to the variability of the experimental conditions in this study,} the original hypothesis adopted from the literature that pre-movement time decreases as the age of children increases was neither confirmed nor ruled out. The current results did not display a clear correlation between pre-movement times and children's ages. Still, the observations revealed that the age of the children does substantially impact the pre-movement phase of the evacuation process. \par
Furthermore, the evacuation behaviour of both staff members and preschool children in our study reflected the daily routines in the specific nursery schools. What is more, children tended to respect well-known, learned rules under all circumstances (e.g., children insisted on finishing their snacks or drinking before leaving their classrooms). This behaviour was generally observed for children in different locations in their buildings carrying out various activities when the alarms went off. Besides following familiar rules of conduct, children also tended to use familiar exits. If the instructions \quotes{go to the door} or \quotes{go outside} were given by staff members, without specification of which exit from the classroom was meant, children automatically chose the main door or the door used in daily practice. Similarly, an exit route leading out of the building that children were familiar with was preferred by children who moved throughout the corridors without any supervision. Therefore, it may be concluded that the affiliation model (theory of affiliation) of escape behaviour proposed by Sime \cite{sime_affiliative_1983, sime_movement_1985} is also applicable to preschool children. Moreover, it should be assumed that the ability of preschool children to choose an appropriate exit route may be limited. \par 
As a part of the analysis of the evacuation drills performed in this study, the levels of physical support given by staff members to children during the pre-movement phase was analyzed. At this point, the obtained results for all age groups were further compared with the findings available in the literature as reported by Larusdottir et al.~\cite{larusdottir_evacuation_2014}, in which the level of assistance received by children three--six~years of age was divided into three groups: \quotes{No physical assistance}, \quotes{Some physical assistance} (which included all forms of physical help except carrying), and \quotes{Carried} (\resub{see Table~\ref{tab:assistance}}). The results from our study were grouped into four categories (details in Section~\ref{subsec:assistance}), where physical contact between staff members and children (excluding carrying) was seen at two levels: the physical assistance necessary to perform the evacuation process (\quotes{Physical contact needed and hand holding}) and gentle contact, when staff members touched children who clearly did not require any assistance (\quotes{Gentle pushing}). This gentle physical contact was mostly observed when children were being counted or when they were exiting through the door of a classroom, and it likely can be more attributed to psychological effects such as emphasising the active and supportive presence of the adult as a provider of emotional security and authority \cite{pianta_mother-child_1997, ainsworth_attachments_1991, koomen_regulation_2003}. Thus, to compare the results with the data from the literature \cite{larusdottir_evacuation_2014}, the category \quotes{Gentle pushing} can be to the large extent considered as a part of the group \resub{\quotes{No physical assistance}}. Overall, the compared findings are in good agreement and show similar trends. \par              
\begin{table}[hbt!]
    \centering
    \footnotesize
    \begin{tabular}{R{4.0cm}|R{2.5cm}|R{4.0cm}|R{2.5cm}}
    \hline
     \multicolumn{2}{R{6.5cm}|}{\textbf{Larusdottir et al.~\cite{larusdottir_evacuation_2014}}}   & \multicolumn{2}{R{6.5cm}}{\textbf{The current study}} \\
     \hline 
      Assistance & Frequency [\%] & Assistance & Frequency [\%]  \\
      \hline
      No physical assistance & 85.9 & \resub{No physical assistance} & 75.0 \\
      \hline
      \multirow{2}{=}{Some physical assistance} & \multirow{2}{=}{12.3} & Gentle pushing  & 18.7 \\
      \cline{3-4}
      & & Physical contact needed and hand holding & 5.9 \\
      \hline
      Carried & 1.8 & Carried & 0.3 \\
      \hline
      \end{tabular}
    \caption{Comparison of levels of assistance required by children during the pre-movement phase presented in this study and the literature.}
        \label{tab:assistance}
\end{table}

\subsection{\resub{Behavioural} statements}\label{sec:statements}
In engineering applications, understanding and quantifying evacuation behaviour is a challenging and complicated task. For this purpose, the concept of \quotes{behavioural statements} has been proven to be a useful tool for presenting and using the currently available knowledge regarding human behaviour in fire from various research studies in egress modeling \cite{gwynne_guidance_2016,kuligowski_guidance_2017}. Based on the literature review and this study's experimental results, a set of behavioural statements about the evacuation behaviour of preschool children in nursery schools were identified and presented (Table~\ref{tab:statement}). This list of pre-movement behavioural statements is supplemented by general statements linked to the entire evacuation process. \resub{These behavioural statements highlight the key factors that should be addressed and considered in evacuation design. They aim to provide the context and theory that supports our understanding of the behaviour of preschool children to be implemented in given evacuation scenarios (a detailed guide on development of behavioural scenarios can be found in \cite{kuligowski_human_2016, nilsson_selecting_2016}). For further details on the original conceptual model of behavioural statements and guidance for their implementation into behavioural models, the reader is also referred to \cite{kuligowski_guidance_2017}. In the following subsections, further interpretation of the behavioural statements and relevant data provided is proposed for fire safety engineers through the performance elements of pre-movement time, evacuation route usage, and behavioural itineraries. These elements can be explicitly or implicitly calibrated by the user in most evacuation models to represent the impact of behavioural statements \cite{kuligowski_guidance_2017}.} 
\begin{table}[htb!]
    \centering
    \footnotesize
    \begin{tabular}{R{0.4cm}|R{14cm}}
    \hline
    \textbf{\#} & \textbf{behavioural statement} \\
    \hline
    \multicolumn{2}{l}{\textbf{Pre-movement phase of the evacuation process}} \\
    \hline
    1 & Preschool children mostly do not react to a warning signal independently; additional impulse and instructions given by responsible staff members are necessary to trigger the evacuation process. Note: This assumption may vary due to various educational practices employed in nursery schools. \\
    \hline
    2 & Reactions of children and staff members to alarms are affected by their actual locations in the building and activities currently performing at the time. \\
    \hline
    3 & Actions and instructions given by staff members vary according to the age of children under their care. \\
    \hline
    4 & The evacuation behaviour of both preschool children and staff members largely reflects their daily routines and rules. Preschool children tend to follow a sequence of memorized daily activities. Staff members appear to be influenced by the ordinary practices established within the nursery school (e.g., practices for organizing children when leaving a classroom). \\
    \hline
    5 &  The level of assistance required by children is age dependent and highly influenced by the staff-to-child ratio per class. \\
    \hline
    6 & During the pre-movement phase, preschool children tend to use familiar exits; however, their final exit choice is usually determined by staff members. \\
    \hline
   
    \multicolumn{2}{l}{\textbf{General statements}} \\
    \hline
    7 & The evacuation of preschool children in nursery schools can potentially be highly organized and is strongly dependent on the reactions, decisions, actions, and preparedness of responsible staff members. \\
    \hline
    8 & Familiarity with evacuation routes and evacuation procedures appears to play an important role in evacuation efficiency. \\
    \hline
    9 & Evacuation procedures, for both the pre-movement and movement phases, can be improved by regular training. \\
    \hline
    \end{tabular}
    \caption{Overview of the behavioural statements for the pre-movement evacuation behaviour of preschool children in nursery schools.}
    \label{tab:statement}
\end{table}
\subsubsection{Pre-movement time}
When estimating pre-movement times, all behavioural statements summarized in Table~\ref{tab:statement} (\#1--9) can considerably affect (i.e., decrease or increase) the expected pre-movement times. Thus, their positive or negative effects should be individually reflected when a specific evacuation scenario is created. 
It is important to note that responsible engineering application of behavioural statements always requires a combination of relevant data sets and careful consideration of the boundary conditions related to the source of the data (such as data collection and analysis methods used to obtain the data itself). The experimental observations described here confirmed that the different levels of announcement and participants' preparedness for evacuation drills both had a non-negligible impact on the measured pre-movement times. In this context, the results obtained during the unannounced evacuation drills can be seen as the most close to real emergency situations, \resub{see Table~\ref{tab:pre-mov_eng}}. For simplicity, the results from this study on pre-movement times measured for the unannounced evacuation drills (summarized \resub{in Table~\ref{tab:pre-mov_eng}}) are related only to two age groups for children: \quotes{Junior under 4~years} and \quotes{Senior over 4~years} (the latter group includes the results for both Senior and Senior+ children; Mixed classes were excluded). \resub{As noted earlier in this paper, the reported pre-movement times are related to the moment when the responsible staff members considered the children ready to leave the classroom (that is, the first child exited the classroom). Therefore, the pre-movement times for the first children represent time delays for entire classes, and the pre-movement times for the last children provide additional information on the time when the classrooms were completely evacuated.}
\resub{The observations of the evacuation experiments also showed the practical importance of regular training of evacuation procedures in nursery schools (\#9). Evacuation drills are generally recommended as a useful tool for improvement of occupant performance in emergency situations \cite{gwynne_future_2020}. This study contributed to the claim that issues during evacuation drills (and real emergency situations) can be minimised through regular and effective training and that best practices about evacuation drills and fire safety education are needed in nursery schools.}\par
\begin{table}[hbt!]
    \centering
    \footnotesize
    \begin{tabular}{R{2cm}|R{4.5cm}|R{4.5cm}}
    \hline
    \multirow{2}{=}{\textbf{}} &  \multicolumn{2}{R{9cm}}{\textbf{Pre-movement time \newline (mean/min/max/SD) [s] (data points)}} \\
    \cline{2-3}
    & \textbf{Junior under 4~years} & \textbf{Senior over 4~years} \\
    \hline
    First child &  35 / 10 / 59 / 15 (9~classes) & 32 / 15 / 54 / 13 (8~classes) \\
    \hline
    Last child & 49 / 27 / 77 / 16 (9~classes)  & 45 / 27 / 62 / 14 (8~classes)  \\
    \hline
    \end{tabular}
    \caption{Pre-movement times measured in the unannounced experimental evacuation drills with children divided into the two age subgroups (Mixed classes excluded).}
        \label{tab:pre-mov_eng}
\end{table}
\subsubsection{Evacuation route usage}
As the experimental observations showed, the choice of an evacuation route during the pre-movement phase of the drills  was made by the responsible staff members. It can be assumed that the staff's decision-making processes in nursery schools are influenced by training (e.g., receiving regular fire protection education), knowledge of evacuation procedures, and good familiarity with the school's enclosure and available evacuation routes. On the other hand, situations may occur when children are not fully under the supervision of staff members or where children receive ambiguous or incomplete instructions from staff. Subsequently, children may choose an escape route (e.g., an exit from a classroom, a direction in a corridor) independently, without guidance from supervisory staff members. In such cases, it can be assumed that children prefer to use familiar exits and escape routes (see behavioural statement \#6 in Table~\ref{tab:statement}). Hence, additional delays in the evacuation process may emerge when the escaping children are found by staff members and redirected to another exit. \par 
\subsubsection{Behavioural itineraries}
To represent evacuation behaviour more specifically, delays and activities which are not directly associated with movement to a place of safety can be assigned to evacuees in more advanced evacuation models \cite{kuligowski_guidance_2017}. Considering the specific conditions of evacuation in nursery schools accommodating preschool age children, several behavioural itineraries can be highlighted.   
During the pre-movement phase of the evacuation process, the interpretation of the warning signals is mostly the responsibility of staff members; children's reactions typically follow the explanations or instructions given by adults (\#1 in Table~\ref{tab:statement}). Thus, the performance of risk identification and risk assessment (i.e., searching for more information) can be expected to be performed by the staff members present. After the decision to escape a building is made, a sequence of activities performed by staff members may arise, including asking children to form pairs, to hold each other's hands, to form a compact group/line, to take a roll call or count children, to gather up belongings, or to open/unlock an exit door. The type of actions taken may vary according to the ages of the children, the activities being performed when an alarm sounds, the location of the children in the building, or daily class/nursery school routines (see also \#2, \#3, and \#4 in Table~\ref{tab:statement}). Staff members may need to carry out additional tasks when children require special physical or verbal support in order to start the evacuation process. Based on the observations during the evacuation drills in this study, 25\% of children received some kind of physical assistance in the pre-movement phase. The level of assistance provided to children was age dependent and highly influenced by the staff-to-child ratio (\#5 in Table~\ref{tab:statement}). All the above-mentioned factors can considerably affect the pre-movement times, and as such, they should be kept in mind when a detailed evaluation of evacuation behaviour is performed.   \par

\section{Conclusions}
This paper presented new experimental data sets and findings on the pre-movement behaviour of preschool children and specific evacuation conditions observed and analyzed during 15~evacuation drills conducted in 10~participating nursery schools in the Czech Republic. Based on the analysis of video recordings, over \numprint{1800}~data points describing the pre-movement evacuation behaviour of 970~children and 87~staff members were gathered, analyzed, and interpreted in close relation to the research methods employed. The main variables of interest were pre-movement times, interpretation of warning signals, reactions of and activities performed by staff members and children, levels of physical assistance provided to the children, and exiting strategies. \par
\section*{Acknowledgments}
The authors thank all participants involved in the evacuation drills for their efforts and cooperation, which made this research project possible. English language editorial guidance was provided by Dr. Stephanie Krueger. 

\section*{CRediT authorship contribution statement}
\textbf{Hana Najmanov\'a}: Conceptualization, \sep 
Methodology, \sep 
Formal analysis, \sep 
Investigation, \sep 
Resources, \sep 
Data Curation, \sep 
Writing - Original Draft; \sep 
\textbf{Enrico Ronchi}: Supervision, \sep 
Methodology, \sep 
Writing - Review \& Editing

\bibliography{Hana_zotero.bib}

\end{document}